**Designing for Work with Intelligent Entities: A Review of Perspectives**

James E. McCarthy

Sonalysts, Inc.

**Author Note**

Correspondence concerning this article should be addressed to Jim McCarthy, Sonalysts, Inc., 2900 Presidential Drive, Suite 100, Fairborn, OH, 45324 United States. Email: mccarthy@sonalysts.com

**Abstract**

As the power of Artificial Intelligence (AI) continues to advance, there is increased interest in how best to combine AI-based agents with humans to achieve mission effectiveness. Three perspectives have emerged. The first stems from more conventional human factors traditions and views these entities as highly capable tools humans can use to accomplish increasingly sophisticated tasks. The second "camp" believes that as the sophistication of these entities increases, it becomes increasingly appropriate to talk about them as "teammates" and use the research on human teams as a foundation for further exploration. The third perspective is emerging and finds both the "tools" and "teammate" metaphors flawed and limiting. This perspective emphasizes "joint activity," "joint cognitive activity," or something similar. In this article, we briefly review these three perspectives.

**Designing for Work with Intelligent Entities:  A Review of Perspectives**

A common goal among Human Factors professionals is to maximize the performance of systems. Increasingly, the systems of interest involve humans interacting with artificially intelligent entities. The entities can be embodied (*i.e.,* robots) or embedded (*i.e.*, software systems or "agents"). In both cases, the entity takes in information about the environment, processes that information in sophisticated ways, and reaches a conclusion on which it can act or share with another stakeholder (*e.g.*, another entity or a human). Across the profession, there is an interest in examining which design metaphors or lenses will lead to the most effective system designs.

Three perspectives have emerged. The first stems from more conventional human factors traditions and views these entities as highly capable tools humans can use to accomplish increasingly sophisticated tasks. The second "camp" believes that as the sophistication of these entities increases, it becomes increasingly appropriate to talk about them as "teammates" and use the research on all-human teams as a foundation for further exploration. The third perspective is emerging and finds both the "tools" and "teammate" metaphors flawed and limiting. This perspective emphasizes "joint activity," "joint cognitive activity," or something similar.

In this discussion, we will briefly review each of these perspectives, their strengths and weaknesses, and some possible paths forward.

**The Case for Tools**

One prominent view is promoted by Ben Shneiderman and others and holds that AI entities are best considered "tools" or "AI-infused supertools" (*cf*. Shneiderman, 2021; Shneiderman & Endsley, 2022; Sarkar, 2023). This perspective stems from the classical human factors literature, especially work on functional allocation, and holds that using language such as "teammate" or "collaborator" diminishes the unique contribution of humans in terms of responsibility, creativity, *etc*., and fails to make appropriate use of the most beneficial and distinctive features of the AI entities (Shneiderman, 2021). Sarkar (2023, p. 6) puts it this way:

> This article advocates for the metaphors AI is a tool, or AI is an instrument. Viewed in this way, the phrase 'human-AI collaboration' becomes inconsistent with how we typically understand the term collaboration. We wouldn't say carpenter-hammer collaboration, for example, or surgeon-scalpel collaboration, or pianist-piano collaboration.

In advocating for the AI-as-a-tool metaphor, Shneiderman and other proponents of Human-Centered AI advocate for the thoughtful consideration of the division of responsibilities and control between humans and the AI-infused tools they use (Shneiderman, 2020). This perspective expands Shneiderman's well-recognized ten levels of autonomy to create a two-dimensional taxonomy of control. In most but not all cases, the goal is to develop systems with high levels of both *system automation* and *human control* (Shneiderman, 2020). The taxonomy is reproduced in Figure 1.

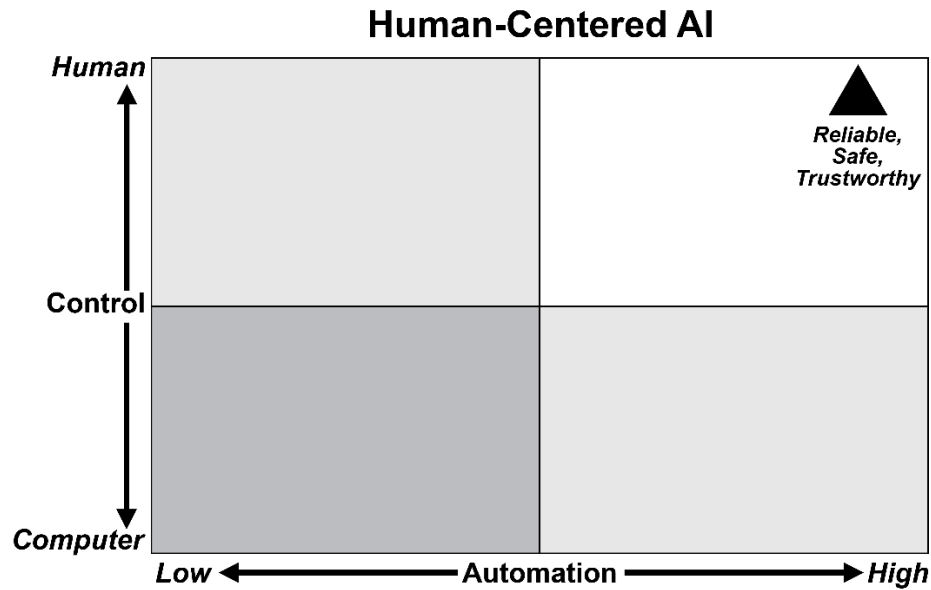

Figure 1: Note: Two-Dimensional Outcome Taxonomy. Adapted from "Human-centered artificial intelligence: Reliable, safe & trustworthy" by B. Shneiderman, 2020, *International Journal of Human–Computer Interaction*, *36*(6), 495-504.

In general, the concept of operations that underlies this framework is captured by the following phrase that expresses the role of the human using the AI-infused supertool: "preview first, select and initiate, then manage execution" (Shneiderman & Endsley, 2022). The goal is to (1) ensure that humans are firmly in the loop by putting them in control of when/whether/which action(s) the supertool completes, and (2) "give human users powerful, but predictable and controllable devices for which they can take responsibility" (Shneiderman, 2021).

While it has many adherents, this entity-as-tool perspective is not the majority view (Shneiderman, 2023b). Those who criticize the perspective primarily do so because they feel it conflates the phrases "entities as teammates of humans" and "entities as human teammates." The former connotes entities that co-operate with humans on interdependent problems; the latter suggests tools that are as capable as humans. These are very different notions, but proponents of the tool metaphor often merge them. To illustrate this point, consider the following passages:

- "The metaphor of 'teammate' seems to encourage designers to emulate human abilities, rather than take advantage of the distinctive capabilities of computers." (Shneiderman, 2021).
- "I became more confident that supporting human performance rather than replacing it was the right idea" (Shneiderman, 2023b).
- "For AI to be a collaborator, we must ask: does it behave in all important respects as a human collaborator does?" (Sarkar, 2023, p. 6).

As we summarize below, other researchers feel that an entity does not need to be capable of perfect performance, demonstrate human-like performance, or even have human DNA to be considered a teammate.

Another set of criticisms that researchers offer to challenge the tool metaphor focuses on the use cases that its proponents employ to illustrate and test the entity-as-tool framework. These use cases, they say, are too simplistic to be of much use. Typical use case examples in the literature include digital cameras, GPS devices, and surgical robotics (Shneiderman & Endsley, 2022). Critics frequently point out that these use cases do not represent the complex, cognitively intensive, and parallel types of tasks that the "teammate" metaphor is thought to address.

## The Case for Teammates

Another design metaphor is to consider the intelligent entities as possible teammates. There is no clear leader of this school of thought (*i.e.,* a counterpart to Shneiderman), but Mica Endsley (*e.g.*, Shneiderman & Endsley, 2022) and Nancy Cooke (*e.g.,* Cooke et al., 2023) have spoken eloquently in support of the perspective.

Within the "entities-as-teammates" school, there are two classes. The first class harkens back to the "entities-as-tools" discussion and uses all-human teams as the analog. As we noted above, this perspective is criticized because the current generation of intelligent entities is not as capable as humans along many dimensions. For example, Groom and Nass (2007) strongly argued that these systems cannot serve as teammates, primarily because they lack robust mental models and have limited self-awareness. Further, they set a high bar for the capabilities that entities must possess to qualify for the "teammate" label:

- Share mental models
- Share a common goal
- Subjugate individual needs for group needs
- View interdependence as a positive
- Know and fulfill their roles
- Trust each other

Others have suggested a slightly lower bar. Some believe that if (1) the entities use AI or similar technologies to make decisions and/or take actions within a specified domain, (2) they can respond to novel performance challenges (McNeese et al., 2018), and (3) are capable of coordination and cooperation (Chiou & Lee, 2021), we can begin to view the system as an autonomous teammate that is capable of not only independent but interdependent action.

Further, many researchers have convincingly noted that while AI-based entities may not yet have the level of sophistication, it is not an inherent attribute of their nature; there is no reason to assume that improved capabilities along each of the "high bar" requirements proposed by Groom and Nass (2007) will not be forthcoming (*e.g.,* Phillips, Ososky, Grove, & Jentsch, 2011; Phillips, Rivera, & Jentsch, 2013; Warta, Kapalo, Best, & Fiore, 2016; Beans, 2018; Gervits, Thurston, Thielstrom, Fong, Pham, & Scheutz, 2020; Seeber, Bittner, Briggs, De Vreede, De Vreede, Elkins, A., ... & Söllner, 2020; Chiou & Lee, 2021).

The second class within the "entities-as-teammates" school recognizes some of the deficiencies of the human analog and, therefore, suggests that human-animal teams are a more useful model (Kapalo, Phillips, & Fiore, 2016; Phillips, Schaefer, Billings, Jentch, & Hancock, 2016). They note that humans and animals have a long history of working as functional and interdependent teams in natural settings (*e.g.,* hunting) and decidedly unnatural settings (*e.g.,* bomb, drug, or disease detection; search-and-rescue). Further, they emphasize that the strengths and limitations of the animal members of the team (*e.g.,* strength, emotions, cognition) serve as suitable analogs for the strengths and limitations of AI entities (Phillips, Schaefer, Billings, Jentch, & Hancock, 2016):

Finally, it is important to note that human-animal team analogs of interdependence teaming illustrate that full human emulation is unnecessary to facilitate meaningful human-AI teamwork. In fact, researchers suggested that full human emulation in human-robot interaction is a somewhat misguided approach, as it ignores the skills that robots can uniquely contribute to a team.

Proponents of the "entities-as-teammates" perspective often point to the heuristic value of this viewpoint. For example, notions such as "mental models" that have their foundation in the study of all-human teams can lead researchers to ask what elements of a mental model might be useful within AI entities. Further, they may also examine the functioning of human-canine (or other animal) teams and explore whether dogs (or other animals) possess anything like a mental model, what the capabilities of the analog might be, and how it might develop. Table 1 illustrates some of the key research questions that emerge from the joint consideration of Human-AI, Human-Animal, and Human-Human teams.

**Table 1**

**Research Questions Stemming from Entities-as-Teammates Perspective**

| Kapalo, Phillips, & Fiore, 2016 | Phillips, Schaefer, Billings, Jentch, & Hancock, 2016 | Cooke et al., 2023 |
|---|---|---|
| • What cues lead to the perception of social presence?<br>• How do team members signal intention?<br>• How do team members interpret intentionality cues?<br>• How are signaling and other non-verbal cues used to support communication?<br>• What level of trust is assumed, and what triggers those assumptions?<br>• How is trust developed or diminished over time?<br>• What is the role of co-learning in the development of trust? | • How can we shape the mental model(s) of human members of teams?<br>• How can we optimize the role/functions of AI entities within hybrid teams?<br>• How are signaling and other non-verbal cues used to support communication?<br>• What level of trust is assumed, and what triggers those assumptions?<br>• How does trust develop or diminish over time?<br>• What is the role of co-learning in the development of trust?<br>• What levels/types of interdependence can be supported in hybrid teams? | • Do teammates have to be humans?<br>• Can role differentiation be used to guide the design of hybrid teams?<br>• Do teams, by definition, need a shared goal or shared identity?<br>• Do all team members need the same perception/understanding of hierarchy or authority differentiation?<br>• Can teammate interdependence vary by degree?<br>• Can interdependencies among teammates vary over time? |

**Alternative Perspectives**

A third perspective is emerging that recognizes that both the "tools" and the "teammates" schools of thought have deficiencies. They note that the "teammate school" sometimes blurs the line between human teammates and non-human cooperating entities in ways that are not helpful. They also point out that the "tool school" often fails to recognize the cognitive nature of the AI entities and the resultant cognitive impact they have on human actors. They believe it is more helpful to refer to these arrangements as "joint activities" or "joint cognitive activities" and avoid both labels. Rather than serving as a design metaphor *per se*, this approach represents a collection of methods and theoretical perspectives that explore how team components (humans, tools, and agents with varying levels of sophistication) can work together to maximize performance (Cochran & Rayo, 2023). Within this perspective, the focus is not on the human members of the team nor the automated portions (whether they are considered tools or teammates); instead, the unit of analysis is the team itself and the critical variable is the team's performance.

Although this perspective has gained increased attention in recent years (*cf.* Xu & Dainoff, 2023; Xu & Gao, 2023; Cochran & Rayo, 2023), it actually dates back four decades (*e.g.*, Woods, 1985). Woods (1985) defined cognitive systems as those that are goal-directed and that use knowledge about themselves and the environment to plan, monitor, and modify their actions in pursuit of those goals. Using this definition, both humans and AI entities can be considered cognitive systems. In recognition of this, Woods (1985) and others argued for a new design language that extended the principles of Human-Computer Interaction (HCI) to account for these "joint cognitive systems."

Hoc (2000) revisited this notion and referred to it as Human-Machine Cooperation (HMC). For Hoc, HMC involved adding to the "know-how" of AI entities by providing "some know-how-to-cooperate," focusing specifically on the cognitive aspects of cooperation.

Michael Rayo and his colleagues (*e.g.,* Rayo, 2017; Morey, Rayo, & Li, 2022) have focused on building on the foundations established by User-Centered Design (UCD) and Cognitive Systems Engineering (CSE) to create a multidisciplinary approach to maximizing the effectiveness of joint activity teams. For example, the team has explored how to structure teams and design activities to maximize the effectiveness of an early warning alarm system and has defined a "symbiotic design process" based on their research on alarm systems (Li *et al*., 2021, p. 211). They have also worked to create novel analysis methods, such as Abstraction Networks (Tewani *et al.,* 2023). Further, they have developed novel evaluation metrics such as Joint Activity Testing (JAT), Joint Activity Monitoring (JAM), and Joint Activity Test Analysis (JATA) that are collectively based on Joint Performance Graphs (JPGs). These testing, monitoring, and evaluating techniques allow researchers to characterize the value of various team compositions in terms of overall performance, brittleness, and extensibility (*e.g.,* Morey *et al*., 2020; Morey *et al.,* 2022).

Similarly, Xu and Gao (2023) introduced the term Human-AI Joint Cognitive Systems (HAIJCS) to extend the foundations established by HCI. Xu and his colleagues used this term to recognize situations in which all the involved entities (human or not) can be considered cognitive systems and within which the goal is to design systems that employ "the complementary advantages of AI and human intelligence to produce a more powerful intelligence form:  human-machine hybrid intelligence."

As noted above, the "joint cognitive activity" school aims to use the strengths of both humans and the AI entities to produce a result that is better than what either is capable of producing by itself. Woods (1985) put it this way:

The human and machine elements contain partial and overlapping expertise that, if integrated, can result in better joint system performance than is possible by either component alone.

While this is a laudable goal, there have been few measurable examples of it occurring. That is beginning to change. For instance, Duros et al. (2019) recently reported on a study in which humans and AI entities cooperated to identify chemical compounds with certain properties. Working independently, human chemists were right approximately 66% of the time. Similarly, the machine intelligence systems were right about 72% of the time. However, when human intuition was used to guide the data-gathering efforts of the intelligent entities, the accuracy increased to approximately 76%. While not yet definitive, the work of Michael Rayo, Dane Morey, and their colleagues is promising. It suggests that developers can use the techniques this research team espouses to increase resilience, maximize joint performance, and enhance the extensibility of socio-technical teams.

## Discussion

In considering the various perspectives reviewed above. I have reached a few tentative conclusions that I share here to prompt more discussion.

**First, I find myself adopting the team performance focus and analysis framework proposed by the joint activity community**. The work of Morey, Rayo, & Li (2022) and others provides a valuable framework for assessing the performance of automation-enhanced systems. In particular, they emphasize continuous measurement of high-validity challenges and meaningful measures of team processes and outcomes. Proponents of this approach also focus on ongoing and non-intrusive measurement within deployed systems. These techniques provide an invaluable source of data that expands well beyond what is practical in a system development or research setting. Regardless of the development perspective one adopts, these points of emphasis are likely to be valuable.

**Second, within the joint activity framework, I find the argument for the "teammate" design metaphor particularly compelling**. Humans have roughly 200,000 years of experience working in teams (Apicella & Silk, 2019). Given this long history, patterns of human teamwork tend to be well-established. Therefore, it is natural to assume that AI entities must be equipped with the same types of "team skills" as their human counterparts to maximize their effectiveness as members of a Human-AI team. While agents are limited in this regard right now, I see no reason to suspect that these deficiencies are fundamental to their nature.

**Third, independent of the design metaphor adopted, practitioners would be wise to assume that *users* will view the intelligent entities as teammates.** There is ample evidence that users will anthropomorphize the entities with little or no prompting. Research indicates that individuals tend to ascribe a "mind" and "intentions" to AI entities (both embodied and embedded) with minimal prompting, and that tendency can have positive and negative impacts on how humans interact with these entities (*cf*. Waytz, Heafner, & Epley, 2014; Wises, Metta, & Wykowska, 2017). For example, Reeves and Nass (1996) demonstrated that people ascribe a social presence to even very basic computer interfaces and react with basic politeness. This tendency has an evolutionary basis and appears to be "wired in" to human cognitive processing (*cf*. Sanfey, Rilling, Aronson, Nystrom, & Cohen, 1993; Wises, Metta, & Wykowska, 2017).

**Fourth, designers can be intentional about designing systems that acknowledge and use this tendency toward personification.** For example, Reeves and Nass (1996) suggested applying Grice's Maxims and other simple rules of etiquette to design user interaction strategies. Miller, Wu, and Chapman (2004) continued this line of thinking by defining a model that could be used to guide when and how to interrupt a user to provide medication reminders. Adherence to these "niceties" has real impacts on the effectiveness of systems with some form of social presence, including influencing the degree to which users trust the recommendations provided (Parasuraman & Miller, 2004).

**Fifth, to support research in this area, it may be useful to pay more attention to how much mind, intentionality, and "teammate-ness" humans attribute to the intelligent entities with which they co-operate.** For example, Waytz, Cacioppo, & Epley (2010) developed a measure of an individual's tendency to anthropomorphize and demonstrated that it was a relatively stable trait. Further, the researchers demonstrated that higher levels of this trait increased the likelihood that participants would:

- Ascribe secondary emotions (*e.g*., shame or optimism) to a non-human entity
- Express concern for a non-human entity's well-being
- Demonstrate increased trust in a non-human entity

A related measure was developed by Wynne and Lyons (2018; 2019). Wynne and Lyons designed their Autonomous Agent Teammate-likeness (AAT) measure to use five-point Likert scale items to assess a given entity along six dimensions:

1. Perceived agentic capability
2. Perceived benevolence/altruistic intent
3. Perceived task interdependence

4. Perceived relationship-oriented behavior

5. Perceived richness of communication

6. Perceived presence of a mind and a shared mental model

It may be helpful to apply the AAT measure in a research setting that adjusts the features of an AI entity in ways similar to those used in the previously mentioned anthropomorphism/theory of mind research. This work could refine the AAT measure and explore its interaction with the anthropomorphizing trait measure developed by Waytz, Cacioppo, & Epley (2010).

A fruitful area of research might be to assess the extent to which designers can use agent design to "prime" or shape appropriate mental models in humans. It would be interesting to investigate the degree to which designers can use specific techniques (*e.g.*, adopting an animal form factor *vs.* a human form factor, controlling the apparent politeness of the automated system) to moderate human expectations and behaviors. Could techniques be used to reduce the ascription of a mind and intentionality to an artificial entity and thereby prevent surprises, the loss of trust, and disuse (Kapalo, Phillips, & Fiore, 2016; Phillips, Schaefer, Billings, Jentsch, & Hancock, 2016; McNeese, Flathmann, O'Neill, & Salas, 2023)?  Would that lead to superior system performance or a generalized lack of use?

**Sixth, in addition to considering how HATs are like all-human teams, consider how they are different and use that to guide design decisions and programs of research.** Attending to the ways that Human-AI teams differ from all-human teams is critical. The lessons of the "supertool" or "joint cognitive activity" perspectives should not be minimized. Similarly, we can draw important lessons from the excellent work on human-animal teaming (*e.g.,* the functioning of teams with little or no verbal communication). Opening ourselves to these other perspectives can also allow us to consider whether the ways in which the AI entity differs from a human teammate are essential elements of their nature or temporary differences that may disappear as the technology matures.

DESIGNING FOR WORK WITH INTELLIGENT ENTITIES                                                                  16Kapalo, K. A., Phillips, E., & Fiore, S. M. (2016, September). The Application and Extension of the Human-Animal Team Model to Better Understand Human-Robot Interaction: Recommendations for Further Research. In *Proceedings of the Human Factors and Ergonomics Society Annual Meeting* (Vol. 60, No. 1, pp. 1225-1229). Sage CA: Los Angeles, CA: SAGE Publications.

Li, M., Morey, D. A., & Rayo, M. F. (2021, June). Symbiotic Design Application in Healthcare: Preventing Hospital Acquired Infections. In *Proceedings of the International Symposium on Human Factors and Ergonomics in Health Care* (Vol. 10, No. 1, pp. 211-216). Sage CA: Los Angeles, CA: SAGE Publications.

McNeese, N. J., Flathmann, C., O'Neill, T. A., & Salas, E. (2023). Stepping out of the shadow of human-human teaming: Crafting a unique identity for human-autonomy teams. *Computers in Human Behavior*, *148*, 107874.

McNeese, N. J., Demir, M., Cooke, N. J., & Myers, C. (2018). Teaming with a synthetic teammate: Insights into human-autonomy teaming. *Human Factors, 60*(2), 262-273.

Miller, C. A., Wu, P., & Chapman, M. (2004, January). The Role of "Etiquette" in an Automated Medication Reminder. In *AAAI Technical Report (4)* (pp. 88-96).

Morey, D. A., Della Vella, D., Rayo, M. F., Zelik, D. J., & Murphy, T. B. (2022, September). Joint activity testing: towards a multi-dimensional, high-resolution evaluation method for human-machine teaming. In *Proceedings of the Human Factors and Ergonomics Society Annual Meeting* (Vol. 66, No. 1, pp. 2214-2219). Sage CA: Los Angeles, CA: SAGE Publications.

Morey, D. A., Marquisee, J. M., Gifford, R. C., Fitzgerald, M. C., & Rayo, M. F. (2020, December). Predicting graceful extensibility of human-machine systems: a new analysis method for evaluating extensibility plots to anticipate distributed system performance. In *Proceedings of the Human Factors and Ergonomics Society Annual Meeting* (Vol. 64, No. 1, pp. 313-318). Sage CA: Los Angeles, CA: SAGE Publications.

DESIGNING FOR WORK WITH INTELLIGENT ENTITIES                                                                18